\documentclass[%
 aip,
 apl,
 amsmath,amssymb,
 reprint%
]{revtex4-1}

\usepackage {CJK}

\usepackage{siunitx}

\usepackage{graphicx}
\graphicspath{ {}}
\usepackage{bm}

\usepackage[spanish,swedish,portuges,german, english]{babel}

\usepackage{hyperref}
\hypersetup{
}

\usepackage{tikz, graphicx} 

\newcommand*{\MagentaTriaDown}{
	{\color[rgb]{1,0,1}
	\begin{tikzpicture}
	\filldraw (0,0) -- (0.075,0.15) -- (-0.075, 0.15) -- cycle;
	\end{tikzpicture}
	}}
	
\newcommand*{\BlackTriaUp}{
	{\color[rgb]{0,0,0}
	\begin{tikzpicture}
	\filldraw (0,0) -- (0.15,0.0) -- (0.075, 0.15) -- cycle;
	\end{tikzpicture}
	}}

\newcommand*{\BlueCircle}{
	{\color[rgb]{0,0,1}
	\begin{tikzpicture}
	\filldraw (0,0) circle (2pt);
	\end{tikzpicture}
	}}
	
\newcommand*{\CyanCircle}{
	{\color[rgb]{0,0.8,0.8}
	\begin{tikzpicture}
	\filldraw (0,0) circle (2pt);
	\end{tikzpicture}
	}}

\newcommand*{\CyanRect}{
	{\color[rgb]{0,0.8,0.8}
	\begin{tikzpicture}
	\filldraw (0,0) rectangle (0.15,0.15);
	\end{tikzpicture}
	}}
	
\newcommand*{\GreenDia}{
	{\color[rgb]{0,0.5,0}
	\begin{tikzpicture}
	\filldraw (0,0) -- (0.1,0.1) -- (0.0, 0.2) -- (-0.1, 0.1) -- cycle;
	\end{tikzpicture}
	}}
	
\newcommand*{\RedDia}{
	{\color[rgb]{1,0,0}
	\begin{tikzpicture}
	\filldraw (0,0) -- (0.05,0.1) -- (0.0, 0.2) -- (-0.05, 0.1) -- cycle;
	\end{tikzpicture}
	}}

\begin{document}
\begin {CJK*} {GB} { }


\title[Characterization of low loss microstrip resonators as a building block for circuit QED in a 3D waveguide]{Characterization of low loss microstrip resonators as a building block for circuit QED in a 3D waveguide}

\author{D. Zoepfl}
\author{P. R. Muppalla}
\author{C. M. F. Schneider}

\affiliation{Institute for Quantum Optics and Quantum Information of the Austrian Academy of Sciences, A-6020 Innsbruck, Austria
}
\affiliation{Institute for Experimental Physics, University of Innsbruck, A-6020 Innsbruck, Austria 
}
\author{S. Kasemann}
\author{S. Partel}

\affiliation{Research Centre for Microtechnology, Vorarlberg University of Applied Sciences, A-6850 Dornbirn, Austria}

\author{G. Kirchmair}
\email{gerhard.kirchmair@uibk.ac.at}
\affiliation{Institute for Quantum Optics and Quantum Information of the Austrian Academy of Sciences, A-6020 Innsbruck, Austria
}%
\affiliation{Institute for Experimental Physics, University of Innsbruck, A-6020 Innsbruck, Austria
}

\date{\today}

\begin{abstract}
Here we present the microwave characterization of microstrip resonators, made from aluminum and niobium, inside a 3D microwave waveguide. In the low temperature, low power limit internal quality factors of up to one million were reached. We found a good agreement to models predicting conductive losses and losses to two level systems for increasing temperature. The setup presented here is appealing for testing materials and structures, as it is free of wire bonds and offers a well controlled microwave environment. In combination with transmon qubits, these resonators serve as a building block for a novel circuit QED architecture inside a rectangular waveguide.

\end{abstract}
                             
\maketitle

\onecolumngrid

This article may be downloaded for personal use only. Any other use requires prior permission of the author and AIP Publishing. This article appeared in~\cite{zoepfl2017} and may be found at \href{https://aip.scitation.org/doi/full/10.1063/1.4992070}{\textit{https://aip.scitation.org/doi/full/10.1063/1.4992070}}. 
\bigskip
\bigskip

Microwave resonators are an important building block for circuit QED systems where they are e.g. used for qubit readout~\cite{gambetta_quantum_2008, axline_architecture_2016}, to mediate coupling~\cite{majer_coupling_2007} and for parametric amplifiers~\cite{bergeal_phase-preserving_2010}. All of these applications require low intrinsic losses at low temperatures ($k_B T \ll h f_r$) and single photon drive strength. In this low energy regime, the intrinsic quality factor is often limited by dissipation due to two level systems (TLS)~\cite{pappas_two_2011, gao_experimental_2008}. These defects exist mainly in metal-air, metal-substrate and substrate-air interfaces as well as in bulk dielectrics~\cite{gao_experimental_2008, barends_minimal_2010, geerlings_improving_2012,chu2016}. Two common approaches exist, to improve the intrinsic quality factor of resonators. Either one reduces the sensitivity to these loss mechanisms by reducing the participation ratio~\cite{gao_experimental_2008,reagor_reaching_2013,wang2015} or tries to improve the interfaces by a sophisticated fabrication process~\cite{bruno_reducing_2015, megrant_planar_2012}. Reducing the participation ratio requires to decrease the electric field strength. This is typically done by increasing the size of the resonator~\cite{geerlings_improving_2012} or even implementing the resonator using three dimensional structures~\cite{reagor_reaching_2013}.

\begin{figure}[t]
    \centering
    \includegraphics[width = 3.37in]{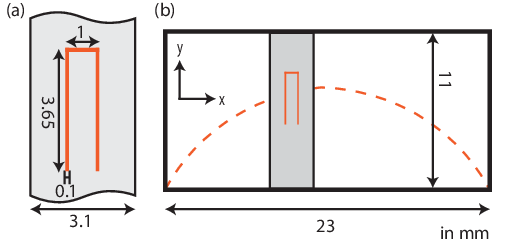}
    \caption{
    \textbf{MSR layout.}(a) Sketch of the MSR on a substrate. (b) Sketch of the cross section of the waveguide with MSR inside. The dashed line indicates the electric field strength of the fundamental mode inside the waveguide.}
    \label{fig:MSR_sketch}
\end{figure}

Our approach, a microstrip resonator (MSR) in a rectangular waveguide (Fig.~\ref{fig:MSR_sketch}), combines the advantages of three dimensional structures with a compact, planar design~\cite{axline_architecture_2016, paik_observation_2011}. The U-shaped MSR (Fig.~\ref{fig:MSR_sketch}(a)) is effectively a capacitively shunted $\lambda/2$ resonator~\cite{pozar_microwave_2012}. The sensitivity to interfaces is reduced, since the majority of the field is spread out over the waveguide, effectively reducing the participation ratio~\cite{wang2015}. Moreover, the U-structure allows a tuneable coupling, by changing the position within the waveguide. Another advantage is, that the waveguide represents a clean and well controlled microwave environment~\cite{pozar_microwave_2012} without lossy seams~\cite{brecht2015} close to the MSR. As the MSR is capacitively coupled to the waveguide, no wirebonds~\cite{wenner_wirebond_2011} or airbridges~\cite{chen_fabrication_2014} are required, which can lead to dissipation or crosstalk.

To assess the performance of different materials, we investigate aluminum and niobium MSRs. The samples were fabricated using standard optical lithography techniques and sputter deposition of the metallic films. Structuring of the metal layer was done using a wet etching process for the aluminum samples and a reactive ion etching (RIE) process for niobium. After completely removing the photoresist, both samples were cleaned in an oxygen plasma.

For microwave transmission measurements, we place the MSR in a rectangular waveguide (Fig.~\ref{fig:MSR_sketch}(b)). The fundamental TE$_{10}$ mode, which has electric field components only along the $y$-axis, is the sole propagating mode at the resonance frequency of the MSR. Its field strength varies along the $x$-axis with a maximum in the center~\cite{pozar_microwave_2012} (dashed line in Fig.~\ref{fig:MSR_sketch}(b)). For the MSR placed off-center, the field strength is different on both legs, which leads to a capacitive coupling to the waveguide. Placed in the exact center of the waveguide, the field strength is equal on both legs of the MSR and the coupling vanishes. Instead of changing the position of the MSR in the waveguide, to change the coupling, we can also fabricate a MSR with legs of different length. To accurately predict the interaction of the MSR with the waveguide we performed simulations of the whole structure using a finite element solver~\cite{see_supp}.

We characterized the MSRs in waveguides fabricated from copper or aluminum. The waveguides were mounted to the baseplate of a dilution refrigerator and cooled down to \SI{20}{mK}. The MSRs were analyzed regarding their resonance frequencies and quality factors by measuring $S_{21}$. We fit the measured data using a circle fit routine~\cite{probst_efficient_2015} which utilizes the complex nature of the $S$-parameter~\cite{see_supp}.

Two sets of measurements were performed. First, we measured the MSRs under variation of input powers, ranging from below the single photon limit to several million photons circulating in the resonator. Second, we stepwise increased the base temperature to \SI{1}{K} and performed measurements at single photon powers.

\begin{figure}[t]
    \centering
    \includegraphics[width = 5.38in]{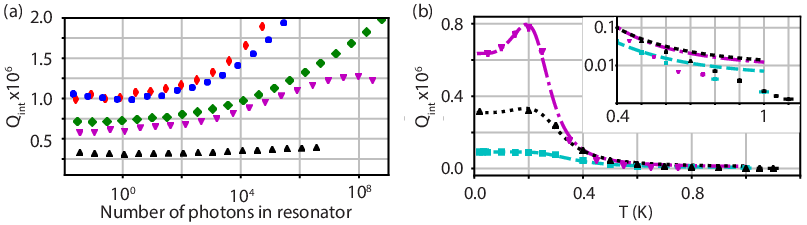}
    \caption{\textbf{Dependence of the internal quality factor on (a) the circulation photon number in the MSR and (b) the temperature (Al MSR).} \protect \GreenDia Nb MSR in Cu waveguide. \protect \BlueCircle \protect \RedDia Nb MSR in Al waveguide. \protect \MagentaTriaDown Al MSR in Al waveguide. \protect \BlackTriaUp \protect \CyanRect Al MSR in Cu waveguide. (a) All but one MSR/waveguide combinations show a clear increase of the internal quality factor with increasing photon number. The only exception is the MSR in the copper waveguide which seems to be limited by other losses. Both Nb MSRs in the aluminum waveguide show an internal quality factor of one million at the single photon limit. The measurements were taken at \SI{20}{mK}. (b) Internal quality factor of Al MSR in dependence of temperature, measured with single photon powers. The data are fitted to a model (dashed lines) combining decreasing TLS related losses (Eq.~\ref{equ:Qi_TLS}) and losses due to an increasing surface resistance with temperature (Eq.~\ref{equ:Qi_Rs}).
    }
    \label{fig:Qi_PowRamp}
\end{figure}

In Fig.~\ref{fig:Qi_PowRamp}(a) we show the dependence of the internal quality factor on the circulating photon number in the MSR. All measurements show a clear trend of an increasing quality factor with the number of photons. This indicates that the MSRs are limited by TLS losses, as they get saturated with increasing drive powers~\cite{pappas_two_2011}. We measure the highest single photon internal quality factor of one million for the two niobium MSRs placed in the aluminum waveguide. For high powers we even measure a $Q_{\text{int}}$ of more than 8 million~\cite{see_supp}. Other experiments, using a more sophisticated fabrication process, report similar internal quality factors for planar NbTiN resonators on deep etched silicon~\cite{bruno_reducing_2015} or for planar aluminum resonators on sapphire~\cite{megrant_planar_2012}. Similar methods and materials might allow us to increase the single photon quality factor of the MSR.

The trend of increasing $Q_{\text{int}}$ is weakest for the aluminum MSR in the copper waveguide, which indicates that this MSR is not limited by TLS. We rather believe that the normal conducting copper waveguide does not shield external fields. Thus vortices might limit the performance of the aluminum MSR in the copper waveguide~\cite{song_microwave_2009}. We do not observe this effect for the niobium MSR, due to its higher critical field~\cite{janjusevic_microwave_2006}. The difference in quality factor of the niobium stripline in the copper and in the aluminum waveguide can be attributed to losses to the copper wall, as suggested by simulations.

We expect two effects on the internal quality factor, when raising the temperature of the MSRs. Approaching the critical temperature leads to a decrease of $Q_{\text{int}}$, due to an increasing surface impedance. Considering a two fluid model~\cite{saito_k._temperature_1999}, the following temperature dependence is found
\begin{equation}
        \frac{1}{Q^{R_s}_{int}} = \frac{A}{T} exp{ \left( - \frac{\Delta}{k_B T} \right) } + \frac{1}{Q_{\text{other}}}.
        \label{equ:Qi_Rs}
\end{equation}
Here $T$ is the temperature, $\Delta$ the superconducting gap at zero temperature, $k_B$ the Boltzmann constant and $A$ a constant. An additional $Q_{\text{other}}$ accounts for other temperature independent losses.  This model is expected to show good agreement until $T_c / 2$.~\cite{gross_festkorperphysik_2014} 

\begin{figure}[t]
    \centering
    \includegraphics[width =5.18in]{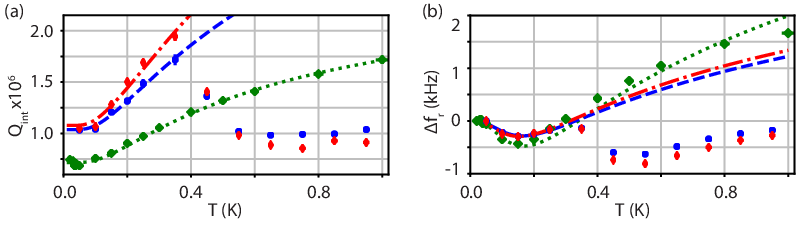}
    \caption{\textbf{Temperature dependence of the Nb MSRs.} \protect \GreenDia Nb MSR in Cu waveguide. \protect \BlueCircle \protect \RedDia Nb MSR in Al waveguide. (a) Internal quality factor of the niobium MSR at single photon powers. The data is fitted (lines) to the TLS model, Eq.~\ref{equ:Qi_TLS}. For the MSR in the aluminum waveguide, this model breaks down around \SI{350}{mK}, when losses of the waveguide walls become dominating. Therefore data points above \SI{350}{mK} are disregarded for the fits. (b) Resonance frequency shift of the MSR at $\approx10^6$ photons. The data is fitted to the model described by Eq.~\ref{equ:fr_TLS}. In the copper waveguide good agreement is observed until around \SI{800}{mK} (data point at \SI{1}{K} is omitted for the fit). In the aluminum waveguide we observe a different behaviour. Above \SI{350}{mK} the observed frequency change is dominated by the waveguide walls (as in (a)). }
    \label{fig:Qi_Nb_Tramp}
\end{figure}

TLS saturate with increasing temperature, which leads to an increase in quality factor~\cite{pappas_two_2011}
\begin{equation}
    \frac{1}{Q_{\text{int}}^{\text{TLS}}} = k \tanh{ \left( \frac{h f_r(T)}{2 k_B T} \right) } + \frac{1}{Q'_{\text{other} }}.
        \label{equ:Qi_TLS}
\end{equation}
Where $k$ is the loss parameter and $h f_r(T)$ represents the energy of the TLS at the resonance frequency of the MSR for a given temperature. The resonance frequency barely changes with temperature (Fig.~\ref{fig:Qi_Nb_Tramp}(b)), which allows us to fix the frequency of the TLS to the resonance frequency of the MSR in the low temperature limit. $Q'_{\text{other}}$ is analogue to Eq.~\ref{equ:Qi_Rs}.

In Fig.~\ref{fig:Qi_PowRamp}(b) we plot the dependence of the internal quality factor of the aluminum MSRs on the base temperature of the dilution cryostat. We fit the data to a combined model~\cite{see_supp} of TLS related losses (Eq.~\ref{equ:Qi_TLS}) and conductive losses (Eq.~\ref{equ:Qi_Rs}). Until about \SI{200}{mK} the MSRs in the copper waveguide show a constant internal quality factor. This gives further evidence that dissipation due to TLS is not the dominant loss mechanism for the aluminum MSRs in the copper waveguide. In the aluminum waveguide we see an increase in $Q_{\text{int}}$ with temperature until \SI{200}{mK}. Thus in this waveguide, TLS related losses most likely limit the quality factor of the MSR. Above \SI{400}{mK} all MSRs show a similar decrease in $Q_{\text{int}}$. This can be attributed to conductive losses, as the critical temperature of aluminum is around \SI{1.19}{K}~\cite{gross_festkorperphysik_2014}. Near the critical temperature, we measure an internal quality factor slightly above 1000. This is close to the results of finite element simulations, which predict an internal quality factor of about 500.

In Fig.~\ref{fig:Qi_Nb_Tramp}(a) we plot the temperature dependence of the internal quality factor of the niobium MSRs. Niobium has a critical temperature of about \SI{9.2}{K}~\cite{gross_festkorperphysik_2014}, hence we do not expect to observe a breakdown of superconductivity. Thus, we only fit the data with the model describing TLS related losses (Eq.~\ref{equ:Qi_TLS}). The behavior of the MSR in the copper waveguide agrees well with predictions from theory throughout the whole measurement range. We observe an increase of $Q_{\text{int}}$ up to \SI{1}{K}. For the MSR in the aluminum waveguide we measure a drop in the internal quality factor at \SI{350}{mK}. In this region we also see the breakdown of superconductivity for the aluminum MSRs (Fig.~\ref{fig:Qi_PowRamp}(b)). This indicates that the breakdown of superconductivity in the waveguide walls is the limiting factor here. For higher temperatures, the internal quality factor remains approximately constant around \SI{1e6}{}. Performing finite element simulations using the finite conductivity of the aluminum (Al5083~\cite{_conductivity_2002}) waveguide wall we found a $Q_{\text{int}}$ of \SI{1.16e6}{}, which is consistent with our measurements.

TLS also lead to a shift in the resonance frequency~\cite{pappas_two_2011}
\begin{equation}
     \Delta f_r (T) = f_r(0) \frac{k}{\pi} \times
         \left( \text{Re} \Psi \left( \frac{1}{2} + \frac{1}{2 \pi i} \frac{h f_r (T)}{k_B T} \right) - \log{ \left( \frac{1}{2 \pi} \frac{h f_r (T)}{k_B T} \right) } \right).
    \label{equ:fr_TLS}
\end{equation}

Here $\Psi$ is the complex digamma function. Fig.~\ref{fig:Qi_Nb_Tramp}(b) shows the frequency shift when increasing the temperature of the cryostat. In contrast to the effect on $Q_{\text{int}}$, off resonant TLS contribute to the frequency shift~\cite{pappas_two_2011}, which makes the resonance frequency independent of power~\cite{see_supp}. The only fit parameter is the combined loss parameter, $k$. For measurements of the Nb MSR in the aluminum waveguide, we observe a drop in the frequency shift above \SI{350}{mK}. We attribute this again to the breakdown of superconductivity in the waveguide wall. Below \SI{350}{mK}, the measurements are in good agreement with the model. In case of the niobium MSR in the copper waveguide, we observe agreement throughout the whole measurement range. 

The values obtained for $k$~\cite{see_supp} by fitting the shift of the resonance frequency are about 10\% to 30\% lower, than fitting the change of the internal quality factor (Fig.~\ref{fig:Qi_Nb_Tramp}(a)). This can be attributed to a non-uniform frequency distribution of TLS~\cite{pappas_two_2011}, which leads to a difference whether $Q_{int}$ or $\Delta f_r$ is considered. The intrinsic quality factor depends on losses to TLS near the resonance frequency, whereas the shift of the resonance frequency depends on a wider frequency spectrum of TLS.

An approximate low power, low temperature limit on $Q_{\text{int}}$ is given by $1/k$. Taking the $k$ value found fitting the change of $Q_{\text{int}}$ gives a 20\% to 30\% higher limit, than found in the measurements. This suggests that the majority of losses happen to TLS, but there is also a second loss mechanism. According to simulations, the internal quality factor of the MSR in the copper waveguide could be limited by the wall conductivity. In the aluminum waveguide it could be attributed to bulk dielectric loss from the high-resistivity silicon, as the loss tangent is not very well known~\cite{chen_fabrication_2014}.

\begin{figure}[t]
    \centering
    \includegraphics[width = 3.37in]{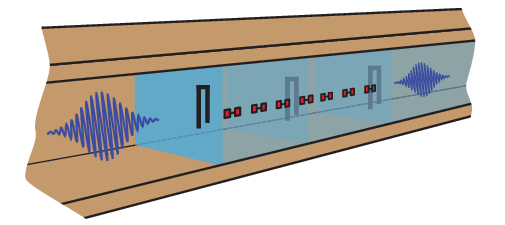}
    \caption{\textbf{MSRs used for analog quantum simulation.} Conceptual schematic of a rectangular waveguide setup using three MSRs (black) as a readout for a chain of transmon qubits (red). The transmon qubits couple capacitively to each other to realize a system for analog quantum simulation of spin chain physics.}
    \label{fig:qubit_msr_wg}
\end{figure}

The presented setup, is an ideal platform for implementing interacting spin systems~\cite{viehmann2013,viehmann2013a,dalmonte2015} where we use the MSR for readout. In Fig.~\ref{fig:qubit_msr_wg} we show a conceptual schematic for simulating spin chain physics. The orientation of the qubits relative to the waveguide allows us to control the coupling of the qubits to the waveguide mode. In Fig.~\ref{fig:qubit_msr_wg} they are oriented along the axis of the waveguide which will lead to a large qubit-qubit interaction but negligible coupling to the waveguide. Three MSRs with different frequencies, all above the waveguide's cutoff, are used to read out selected qubits. Another interesting aspect of this setup is the built-in protection from spontaneous emission due to the Purcell effect, similar to~\cite{reed2010,sete2015} but broadband. Even though the qubit is strongly coupled to the resonator it can not decay through the resonator, as the waveguide acts as a filter if the qubit frequency is below the cutoff.

This platform can also be used to investigate the interplay between short range direct interactions, long range photon mediated interaction via the waveguide~\cite{loo2013} and  dissipative coupling to an open system. It offers a new route to investigate non-equilibrium condensed matter problems and makes use of dissipative state engineering protocols to prepare many-body states and non-equilibrium phases~\cite{diehl2008,cho2011}.

In conclusion, we have presented a design for MSRs with a low interface participation ratio embedded in a rectangular waveguide. The MSRs show single photon intrinsic quality factors of up to one million at \SI{20}{mK}. We find a strong dependence of the internal quality factor on the photon number and the temperature which indicates losses to two level systems. The presented setup is appealing for testing the material of the MSR, the substrate it is patterned on and for validating fabrication processes. The observed quality factors are expected to increase when more complex designs are used, such as suspended structures~\cite{chu2016} or by improving the surface quality through deep reactive ion etching~\cite{bruno_reducing_2015}. Alternatively, switching to sapphire as a substrate is expected to improve quality factors, as interfaces on silicon generally show higher loss than those on sapphire~\cite{chu2016}.

\section*{Supplementary material}
Technical details and further measurement results are shown in the supplementary material.

\newpage
\section*{Acknowledgements}

We want to thank our in-house workshop for the fabrication of the waveguides.

This project has received funding from the European Research Council (ERC) under the European Union's Horizon 2020 research and innovation program (grant agreement n$^{\circ}$ 714235). MP, GK is supported by the Austrian Federal Ministry of Science, Research and Economy (BMWFW). CS is supported by the Austrian Science Fund FWF within the DK-ALM (W1259-N27).

%

\clearpage
\onecolumngrid
\begin{center}
\textbf{\large Supplemental Material: Characterization of low loss microstrip resonators as a building block for circuit QED in a 3D waveguide}
\end{center}
\setcounter{equation}{0}
\setcounter{figure}{0}
\setcounter{table}{0}
\setcounter{page}{1}
\makeatletter
\renewcommand{\theequation}{S\arabic{equation}}
\renewcommand{\theHequation}{S\arabic{equation}}
\renewcommand{\thefigure}{S\arabic{figure}}
\renewcommand{\theHfigure}{S\arabic{figure}}
\renewcommand{\bibnumfmt}[1]{[S#1]}
\renewcommand{\citenumfont}[1]{S#1}
\vspace{0.5 in}

In here technical details and further measurement results are shown, exceeding main paper. We show the full measurement setup. Then we show photographs of a MSR and a full assembled waveguide. The circle fit is discussed, used to analyze the measurements. Also example measurements are presented. Further, the coupling between MSR and waveguide is discussed. Here we show simulation data on the coupling, which were particularly helpful to achieve critically coupled setups. Afterwards, we show measurement results and compare these results to the predictions from simulations. Moreover, we show additional measurements to the main paper. We discuss the resonance frequencies of the MSRs and the highest $Q_{\text{int}}$ measured for the niobium MSR. Moreover, we discuss the shift of the resonance frequency of the aluminum MSR with increasing temperature. Finally, the results of the fits shown in the main paper are given in full detail.

References without leading S relate to the main article.

\section*{Full setup}
\begin{figure}[ht]
    \centering
    \includegraphics[width = 2in]{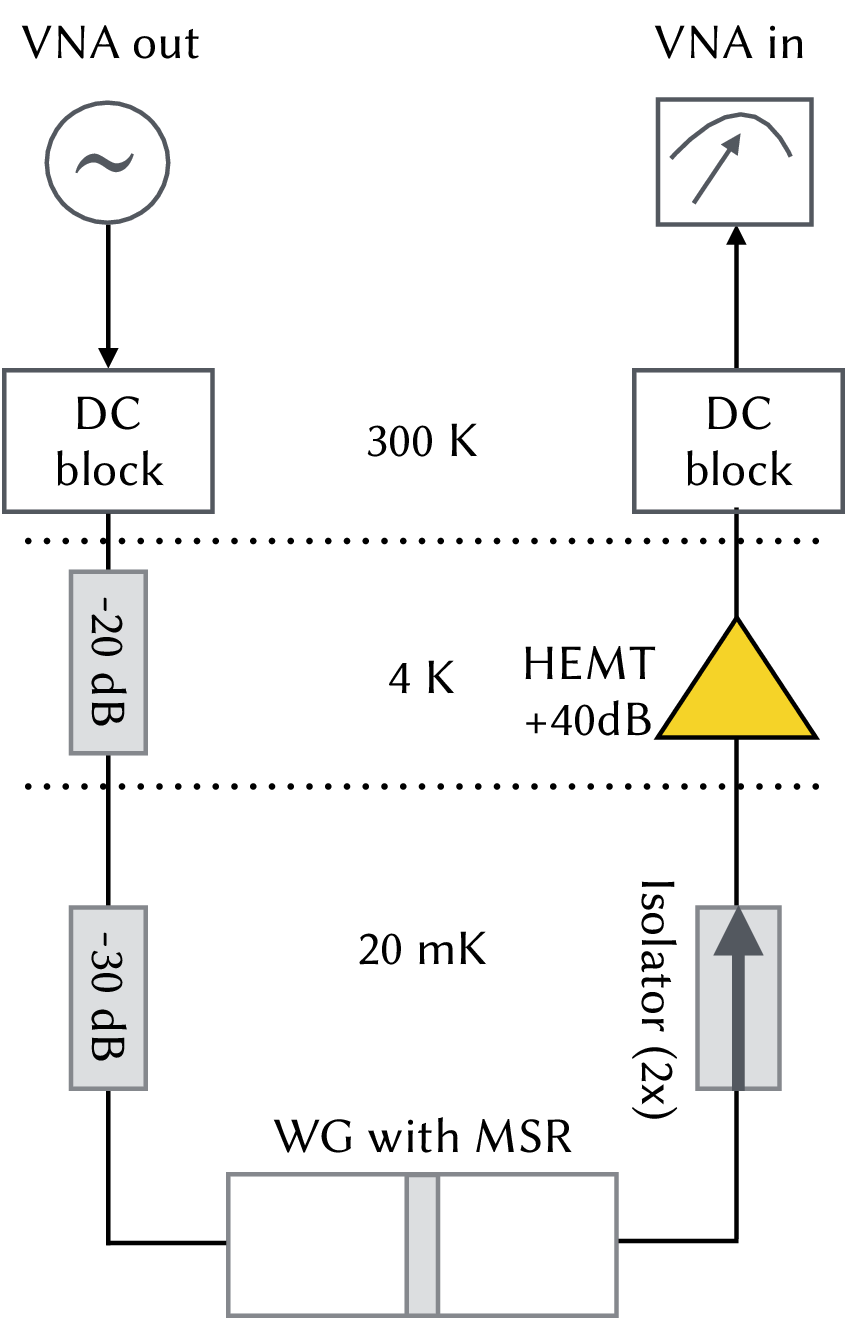}
    \caption{Full measurement setup. Details are explained in the text.}
    \label{fig:mm_setup}
\end{figure}
The full measurement setup is illustrated in Fig.~\ref{fig:mm_setup}. The VNA generates the microwave signal, indicated with 'VNA out', to probe the sample, in this case the MSR. A DC block follows the VNA, which prevents DC currents. The black lines represent microwave lines. The signal enters the cryostat after the DC block and  gets attenuated at \SI{4}{K} by \SI{20}{dB} and further by \SI{30}{dB} at the base plate. The base plate is at a temperature of \SI{20}{mK}. It then enters the waveguide, containing the MSR. The microwave propagates through the waveguide, interacts with the MSR, and leaves it at the other end. The sample is enclosed in a double layer cryoperm shield inside a completely closed copper can. 

At the \SI{4}{K} stage the signal is amplified by \SI{40}{dB}, using a high electron mobility transistor (HEMT) amplifier. Two isolators, which are placed between the HEMT and the waveguide, protect the waveguide from HEMT noise. Another DC block after the HEMT prevents DC currents. Finally the VNA measures the microwave signal. The temperature sensor sits at the base plate of the fridge.

\section*{MSR and waveguide in detail}
In this part we show photographs of the MSR on the silicon substrate, the MSR in the waveguide and the completely assembled waveguide.

\begin{figure}[ht]
    \centering
    \includegraphics[width = 6.06in]{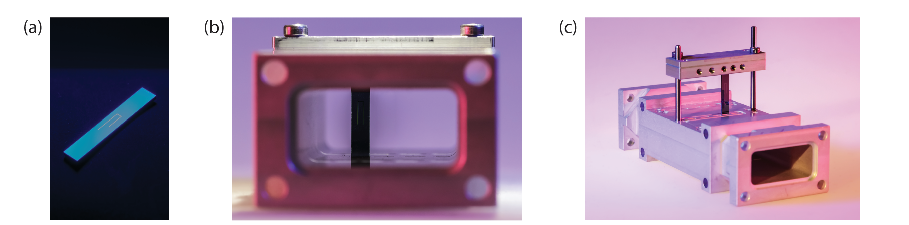}
    \caption{(a) Photograph of a MSR with uneven leg length. (b) Photograph of MSR placed inside the rectangular waveguide. Illustration, including dimensions of the waveguide and the MSR in Fig.~1. (c) Mounting process. The MSR is slid in from the top. Two metal rods are used as guidance.}
    \label{fig:mm_setup_supp}
\end{figure}

In Fig.~\ref{fig:mm_setup_supp} we show a photograph of the MSR (a) and the MSR in the waveguide (b). In (c) the process of mounting the MSR is shown.  We slide the MSR, which is assembled to a holder, from the top into the waveguide. Details about the MSR, including the fabrication, are discussed in the main article. Also the placement of the MSR inside the waveguide is discussed there.

\begin{figure}[ht]
    \centering
    \includegraphics[width = 4in]{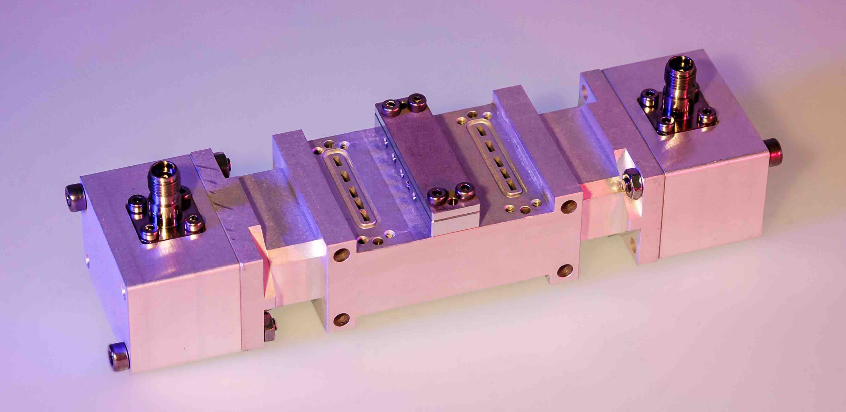}
    \caption{Photograph of totally assembled waveguide. The waveguide we used for the transmission measurement consist of three parts. In the middle section, three rows of samples can be mounted. In this picture only a the middle row is used for a sample.}
    \label{fig:wg_fullAssembly}
\end{figure}

Fig.~\ref{fig:wg_fullAssembly} shows the fully assembled waveguide. The waveguide consists of three parts. At each end there is an identical coupler to receive and launch the microwave signals. The central part contains the samples. In this design no seams are present near the samples. It is possible to probe three samples, each in one of the three slots, which can be easily extended to more samples by using a different central section. One can see the individual slots for each sample, customized to the dimensions of our sample. In contrast to this waveguide, the copper waveguide only allows to probe a single sample at a time.

\section*{Example measurement and circle fit routine}
\begin{figure}[ht]
    \centering
    \includegraphics[width = 0.5\textwidth]{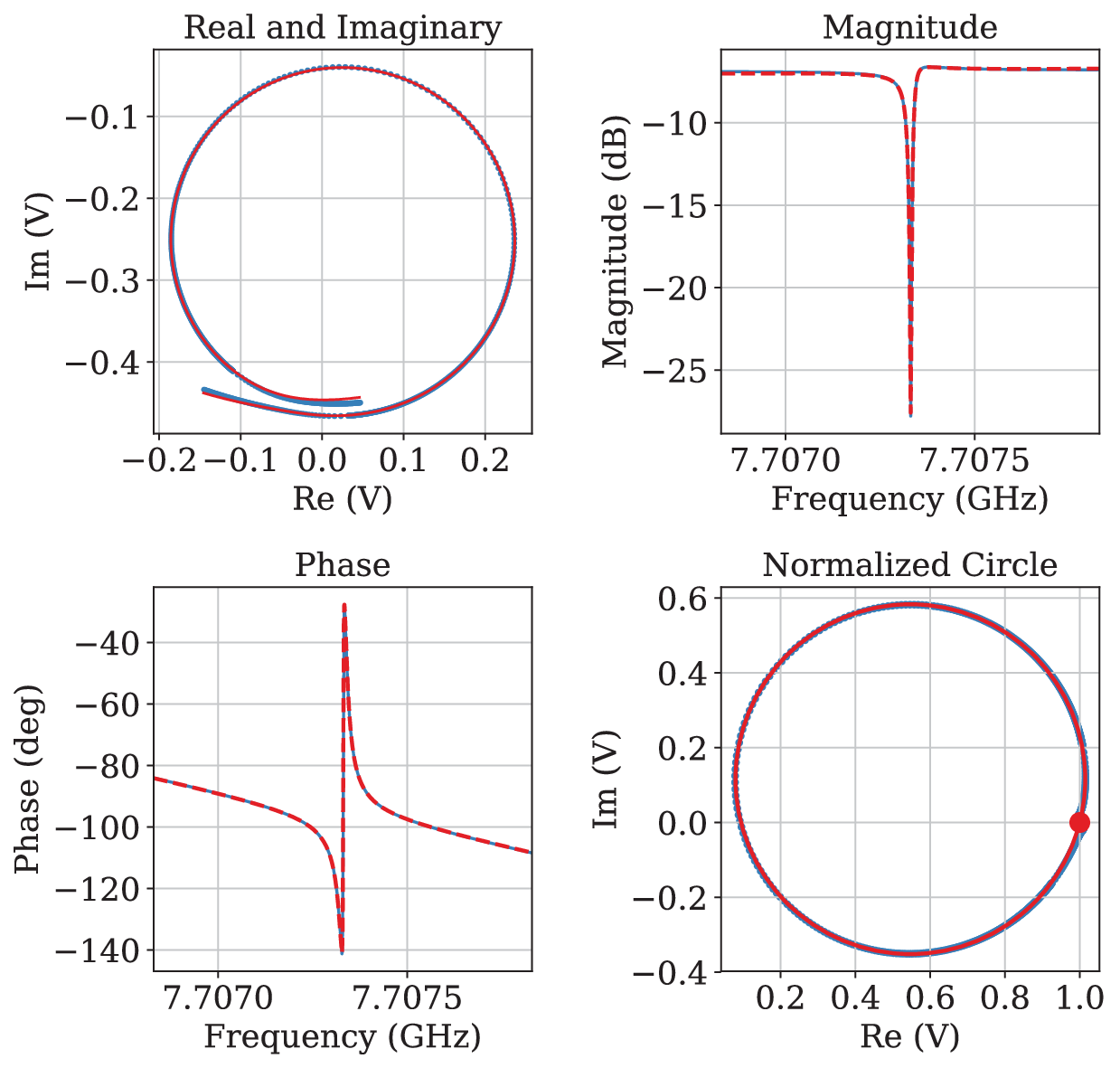}
    \caption{Measurement data of a critically coupled niobium MSR with good signal to noise. Blue is the measured data, red is the fit. The fitting routine gives $Q_c = \SI{4.083(4)e5}{}$ and $Q_{\text{int}} = \SI{3.93(4)e6}{}$. The average photon number of photons circulating in the MSR was \SI{3.4e8}{}.  Top left: Measurement data, imaginary part versus the real part of $S_{21}(f)$, throughout the measured frequency range. Top right: Magnitude of the $S$ parameter versus frequency. Bottom left: Phase versus frequency. The cable delay is seen and the abrupt phase change emerges, where the resonance occurs. Bottom right: Normalized circle (after all effects from the environment are subtracted). The red dot marks the off resonance point exactly opposite to the resonance point.}
    \label{fig:meas_data_lowNoise}
\end{figure}

\begin{figure}[ht]
    \centering
    \includegraphics[width = 0.5\textwidth]{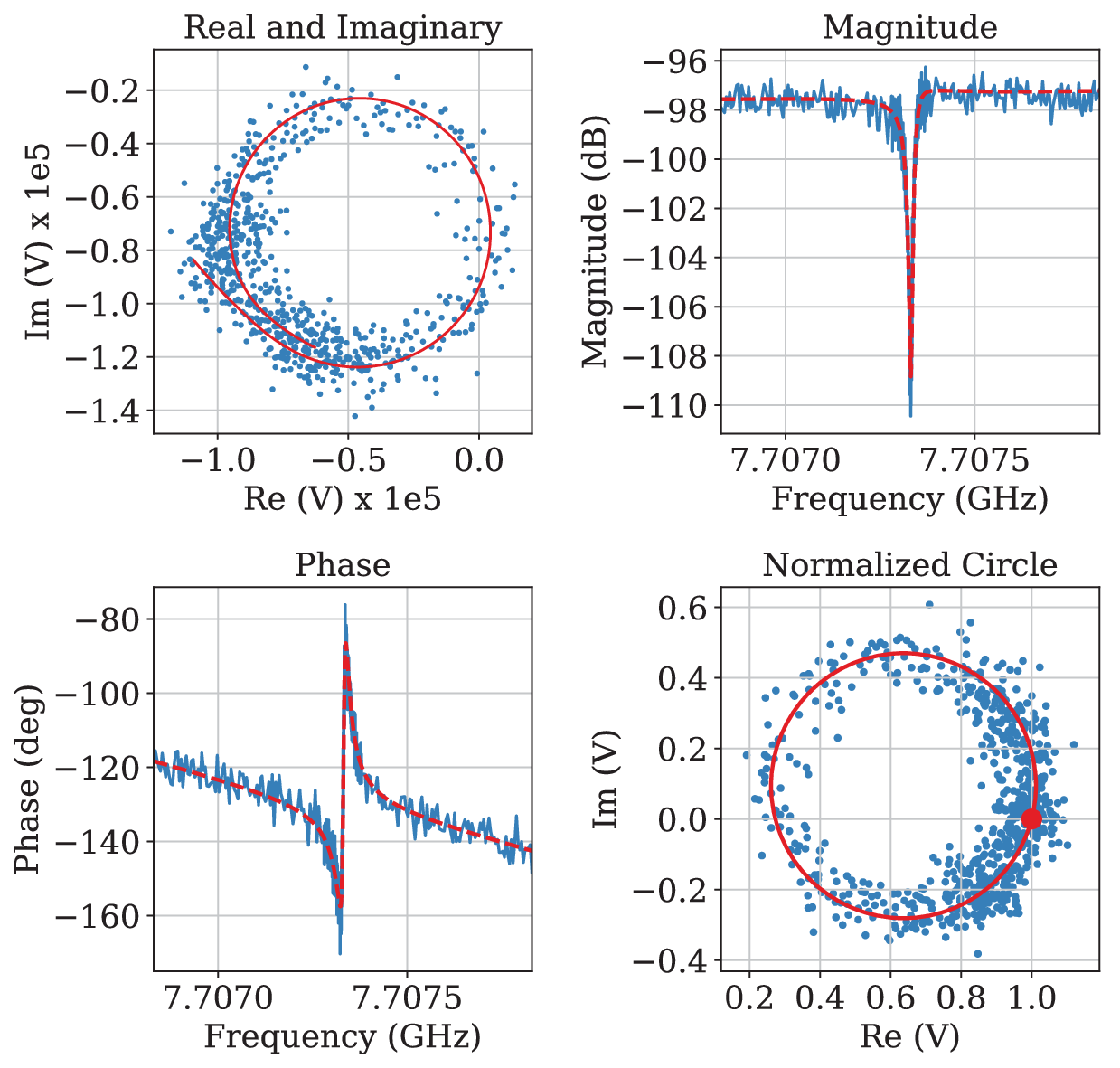}
    \caption{Measurement data of a niobium MSR with low signal to noise near critically coupled. Blue is the measured data, red is the fit. The fitting routine gives $Q_c = \SI{3.97(9)e5}{}$ and $Q_{\text{int}} = \SI{1.06(5)e6}{}$. The average number of photons circulating in the MSR was 0.02. Thus the measurement took several hours. The measured MSR was the same one as in Fig.~\ref{fig:meas_data_lowNoise}.  Top left: Measurement data, imaginary part versus the real part of $S_{21}(f)$, throughout the measured frequency range. Top right: Magnitude of the $S$ parameter versus frequency. Bottom left: Phase versus frequency. The cable delay is seen as a constant slope and the abrupt phase change appears at the resonance frequency. Bottom right: Normalized circle (after all effects from the environment are subtracted). The red dot marks the off resonance point, which is exactly opposite of the resonance point.}
    \label{fig:meas_data_HighNoise}
\end{figure}

The MSR in the waveguide represents a resonator in notch configuration~\cite{S_probst_efficient_2015}. For such a resonator, the $S_{21}$ parameter, which refers to a transmission measurement, follows like~\cite{S_probst_efficient_2015}:
\begin{equation}
     S_{21} (f) =  1 - \frac{Q_l/|Q_c| e^{i \phi_0}}{1 + 2 i Q_l \frac{f - f_r}{f_r}}
     \label{equ:S21_ideal}
\end{equation}
Here $Q_l$ is the total quality factor, $f_r$ is the resonance frequency and $Q_c$ is the coupling quality factor. In here $\phi_0$ accounts for an impedance mismatch in the transmission line before and after the resonator, which makes $Q_c$ a complex number ($Q_c = |Q_c| e^{- i \phi_0}$). The real part of the coupling quality factor determines the decay rate of the resonator, in our case the emission to the waveguide. The physical quantity is the decay rate, $\kappa$, which is inversely proportional to the quality factor~\cite{S_khalil_analysis_2012-1} and therefore the real part is found as: $1/Q_c^{Re} = \text{Re} (1 / Q_c) = \cos{\phi_0} / Q_c$.  Knowing $Q_c^{Re}$ and $Q_l$, the internal quality factor can be obtained, as $1/Q_l = 1/Q_c^{Re} + 1/Q_{\text{int}}$~\cite{S_khalil_analysis_2012-1}. For simplification in all other chapters, $Q_c$ refers to the real part, $Q_c^{Re}$.

Plotting the imaginary versus the real part of $S_{21}$ forms a circle in the complex plane (in case of a resonance within the frequency range).

Equation~\ref{equ:S21_ideal} represents an isolated resonator, not taking effects from the environment into account. Including the environment, which arises by including the whole measurement setup before and after the MSR (Fig.~\ref{fig:mm_setup}), we have to modify Eq.~\ref{equ:S21_ideal} to~\cite{S_probst_efficient_2015}:
\begin{equation}
     S_{21} (f) = (a e^{i \alpha} e^{- 2 \pi i f \tau}) \left(  1 - \frac{Q_l/|Q_c| e^{i \phi_0}}{1 + 2 i Q_l \frac{f - f_r}{f_r}} \right)
     \label{equ:S21_full}
\end{equation}
Here $a$ and $\alpha$ are an additional attenuation and phase shift, independent of frequency. $\tau$ represents the phase delay of the microwave signal over the measurement setup, which has a linear dependence on frequency.

In Fig.~\ref{fig:meas_data_lowNoise} and Fig.~\ref{fig:meas_data_HighNoise} example measurement are shown including the circle fit, Eq.~\ref{equ:S21_full}. Fig.~\ref{fig:meas_data_lowNoise} shows a measurement taken with high power and therefore a good signal to noise. The plot in the top left shows the unmodified measurement data. The imaginary part vs. real part of the $S_{21}$-parameter is plotted with frequency. On the top right and the bottom left, the magnitude and the phase of the measured data is seen. All plots also show the fit. In the bottom right, the data is shown free from the effects of the environment (see equation~\ref{equ:S21_ideal}). Here we also see that the setup is near critically coupled. In Fig.~\ref{fig:meas_data_HighNoise} a measurement below the single photon limit, hence with a low signal to noise, is plotted. The panels show the same information as in Fig.~\ref{fig:meas_data_lowNoise}. In the case of critically coupled setups, trustworthy results are achievable within a feasible measurement time of several hours for low powers.

\section*{Details on coupling between the waveguide and the MSR}

As discussed in the main text, the coupling depends on the position of the MSR in the waveguide along the $x$-axis (Fig.~1). A critically coupled setup is inevitable to get trustworthy results of $Q_{\text{int}}$ and $Q_c$, in particular in the single photon limit, see Fig.~\ref{fig:meas_data_HighNoise}. To accomplish such a setup, simulations were performed. After discussing those, the measurement results are compared to predictions from simulations.

\subsection*{Finite element simulations on the coupling}
We ran simulations on the coupling between the MSR and the waveguide, using a finite element method.

\begin{figure}[ht]
    \centering
    \includegraphics[width = 0.6\textwidth]{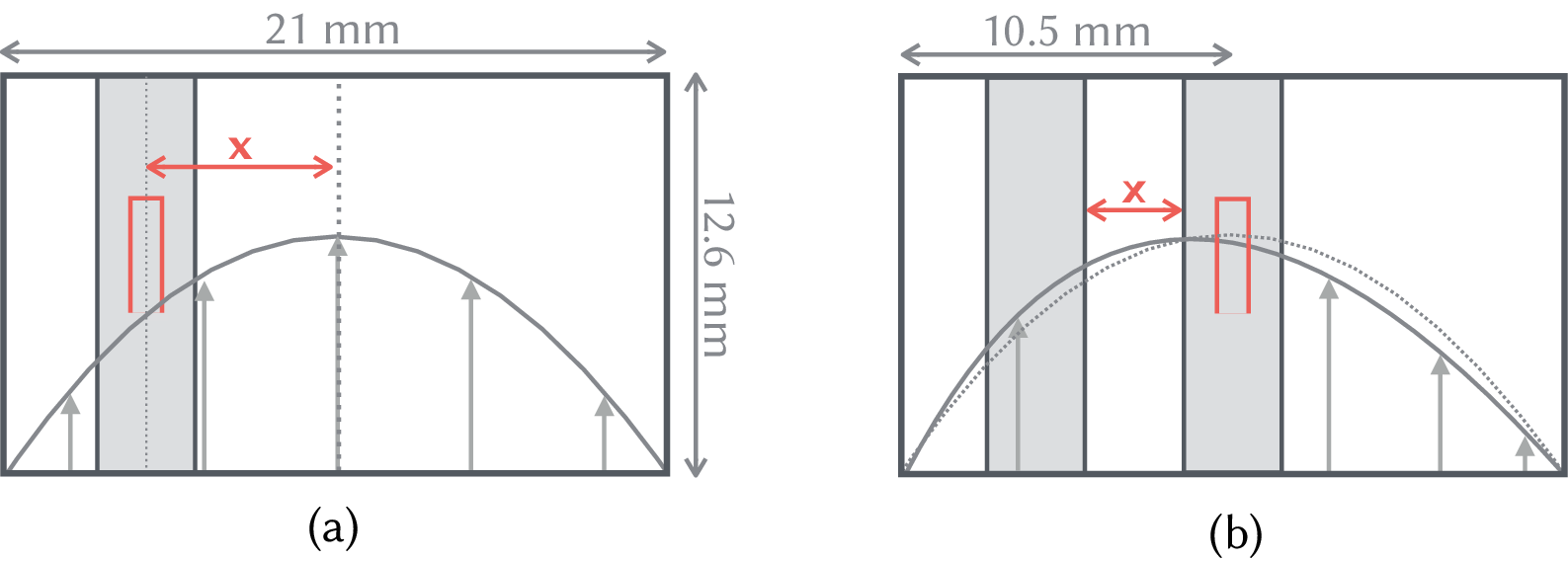}
    \caption{Sketch of the simulated setups. (a) In simulations the MSR was shifted from the center towards the wall. (b) MSR placed in the center with an additional sapphire substrate next to it, which displaces the field inside the waveguide. This leads to an asymmetry of the electric field over the centrally placed MSR, thus a non vanishing coupling. In simulations the substrate was shifted from the MSR towards the wall.}
    \label{fig:sketch_msr_sim}
\end{figure}

\begin{figure}
    \centering
    \includegraphics[width = 0.8\textwidth]{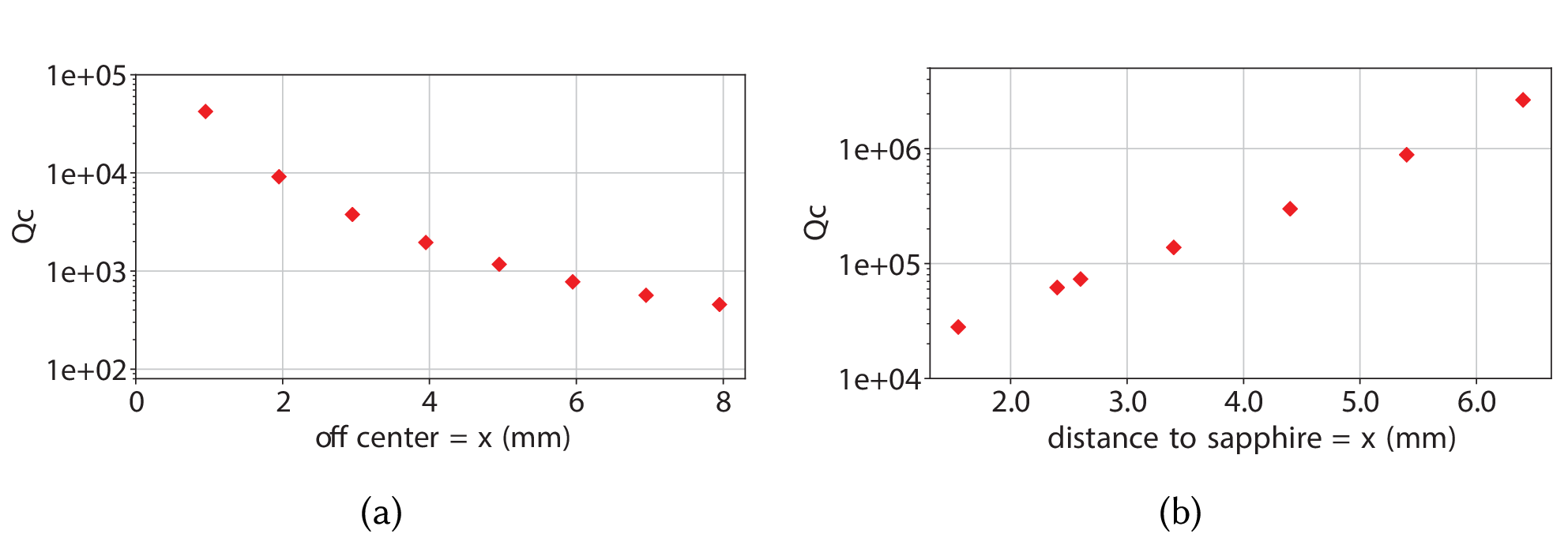}
    \caption{Simulation results of the coupling quality factor for different positions of the MSR in the waveguide. (a) MSR shifted from the center to the waveguide wall (illustrated in Fig.~\ref{fig:sketch_msr_sim}(a)). (b) MSR placed in the center and the empty substrate being shifted away towards the wall (Fig.~\ref{fig:sketch_msr_sim}(b)).}
    \label{fig:sim_msr_pos}
\end{figure}

Fig. ~\ref{fig:sketch_msr_sim} illustrates the two considered cases. At first, the MSR was swept from the center towards the wall. Fig.~\ref{fig:sim_msr_pos}(a) shows the results of the coupling. A critically coupled setup in the single photon limit requires a coupling quality factor on the order of $\SI{1e5}{}$ to $\SI{1e6}{}$, depending on the measured MSR (see Fig.~2). In the available waveguides, there are only discrete slots to place the sample. The first off-centered slot is around \SI{3}{mm} from the center, which leads to a coupling quality factor between \SI{1e3}{} and \SI{1e4}{}, being around two magnitudes below critically coupled.

We decided to use a different approach, with the MSR in the center and an empty sapphire substrate in a neighbouring slot to displace the electric field, due to the higher $\epsilon_r$ of sapphire. This is illustrated in Fig.~\ref{fig:sketch_msr_sim}(b). We ran simulations with the MSR placed in the center and the empty substrate being shifted towards the wall (Fig.~\ref{fig:sim_msr_pos}(b)). The substrate being one slot off center (neighbouring the MSR) leads to a coupling of around \SI{1e5}{}. For two slots off center we can reach the desired coupling quality factor of around \SI{1e6}{}. We should remark at this point, that for such high coupling quality factors, effects like the MSR having a slightly asymmetric leg length, or being placed off center on the chip or placed entirely off center, can have a big impact on the coupling. For instance simulations showed, that a displacement of \SI{0.2}{mm} off-center, can lead to a factor of 4 in the coupling quality factor.

\subsection*{Measurement results}
\begin{figure}[ht]
    \centering
    \includegraphics[width = 0.5\textwidth]{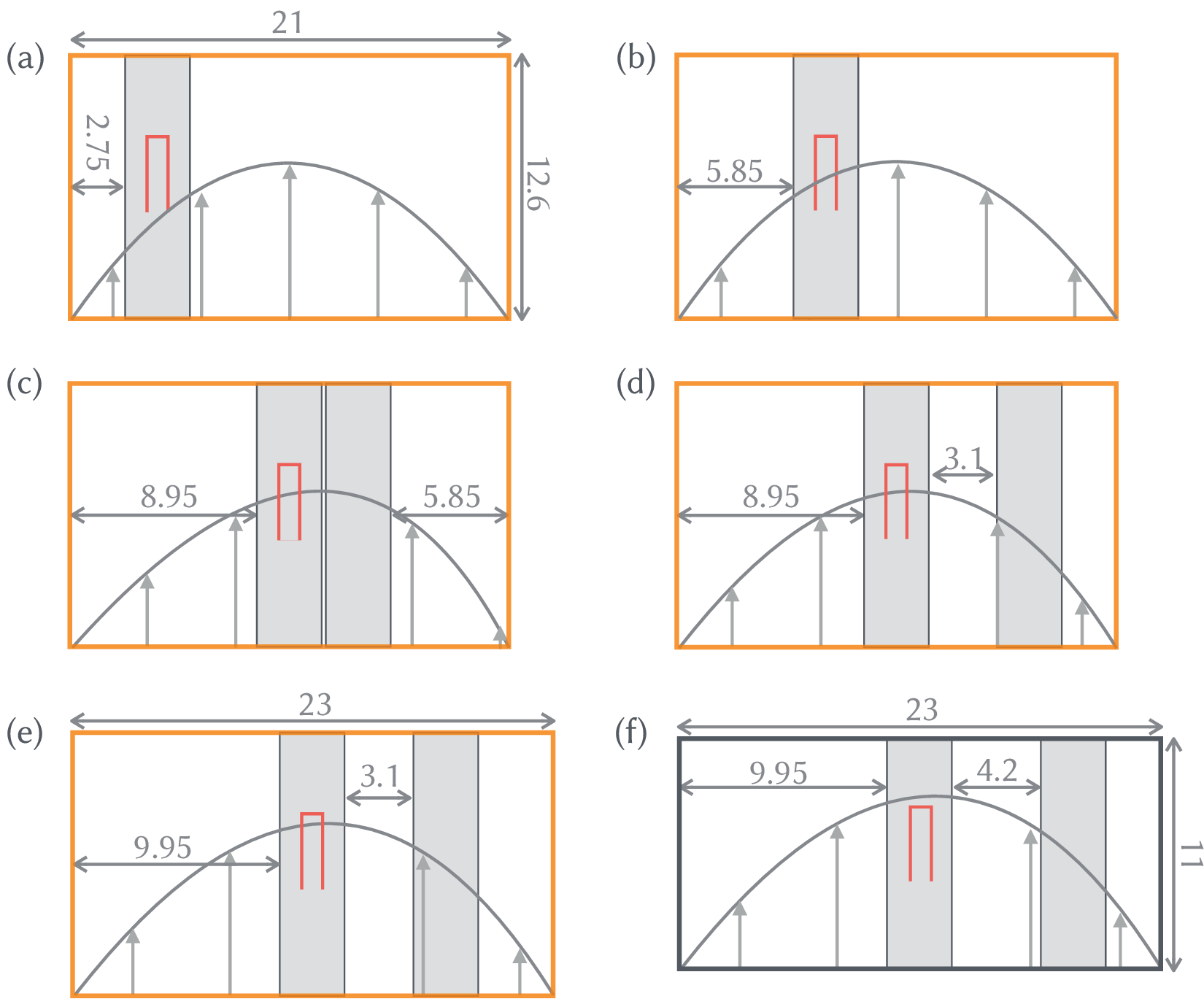}
    \caption{Measured configurations of MSRs in the waveguide. The orange border refers to a waveguide made from copper, the gray border to one made from aluminum. In the copper waveguide, one sample was tested at each time, whereas in the aluminum waveguide three samples were tested at once, refereed to as (f1)-(f3). The dimensions of the different waveguides are stated. All dimension in mm.}
    \label{fig:wg_setups}
\end{figure}

Fig.~\ref{fig:wg_setups} illustrates the actually measured configurations. In the copper waveguides (a) - (e), one sample was measured each time, in the aluminum waveguide, three samples could be measured at once, labelled (f1)-(f3) in the following. The three samples in the aluminum waveguide were put along the propagation direction, all in the same configuration. As they had to be apart in frequency, one MSR had longer legs (f3), leading to a nominally lower resonance frequency of around \SI{0.5}{GHz}. We backed this MSR, as well as a second one (f2), having a resonance frequency of nominally \SI{8}{GHz},  with an empty silicon substrate. This reduces the resonance frequency, due to the higher effective dielectric constant.

\begin{figure}[ht]
    \centering
    \includegraphics[width = 0.45\textwidth]{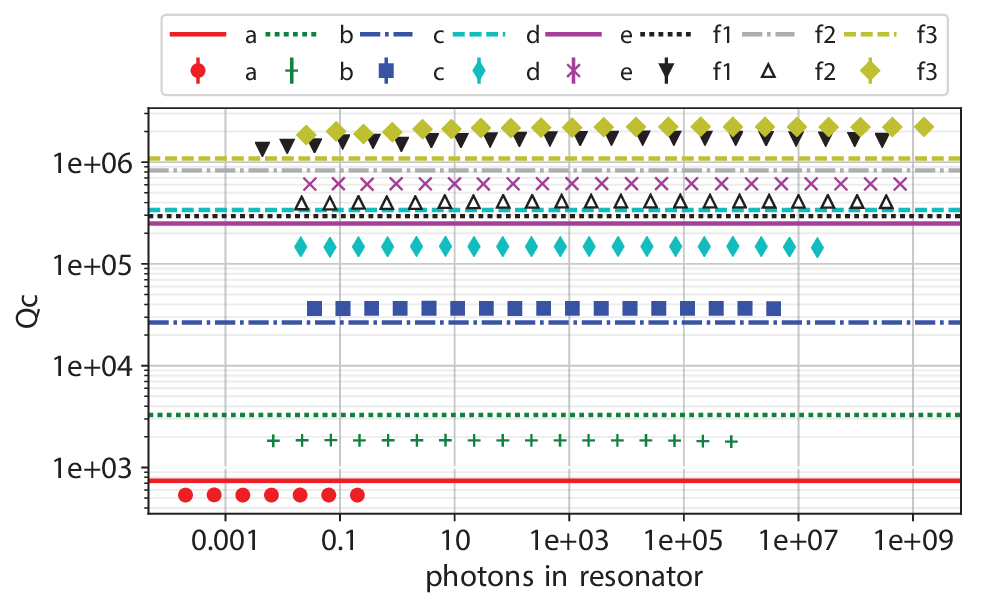}
    \caption{Coupling quality factor for different configurations (see Fig.~\ref{fig:wg_setups}). The lines depict the simulated data, the points are the measurements.}
    \label{fig:Qc_meas}
\end{figure}

The results are plotted in Fig.~\ref{fig:Qc_meas}. We observe overall agreement between the measurements and the simulation data. The weak dependence on the number of photons agrees well with the expected power-independence of the coupling. The closer the MSR is to the wall, the lower the coupling quality factor (a), (b), inline with simulations. With the additional empty substrate, the quality factors follow the predictions from simulations (c), (d). For the substrate further away from the MSR, (d)-(f), highest coupling quality factors are observed. We attribute the difference in the coupling between configurations (d), (e) and (f1), which should be similar (the only nominal difference is the waveguide width) to a slight displacement of the MSR as discussed before. The quality factors of (f2) and (f3) are higher, as they are already backed with an empty substrate. Thus the relative influence of the neighbouring substrate is reduced, leading to a higher $Q_c$.

\section*{Measurements results exceeding the main paper}

\subsection*{Resonance frequency in dependence of photon number in the MSR}

\begin{figure}[ht]
    \centering
    \includegraphics[width = 0.8\textwidth]{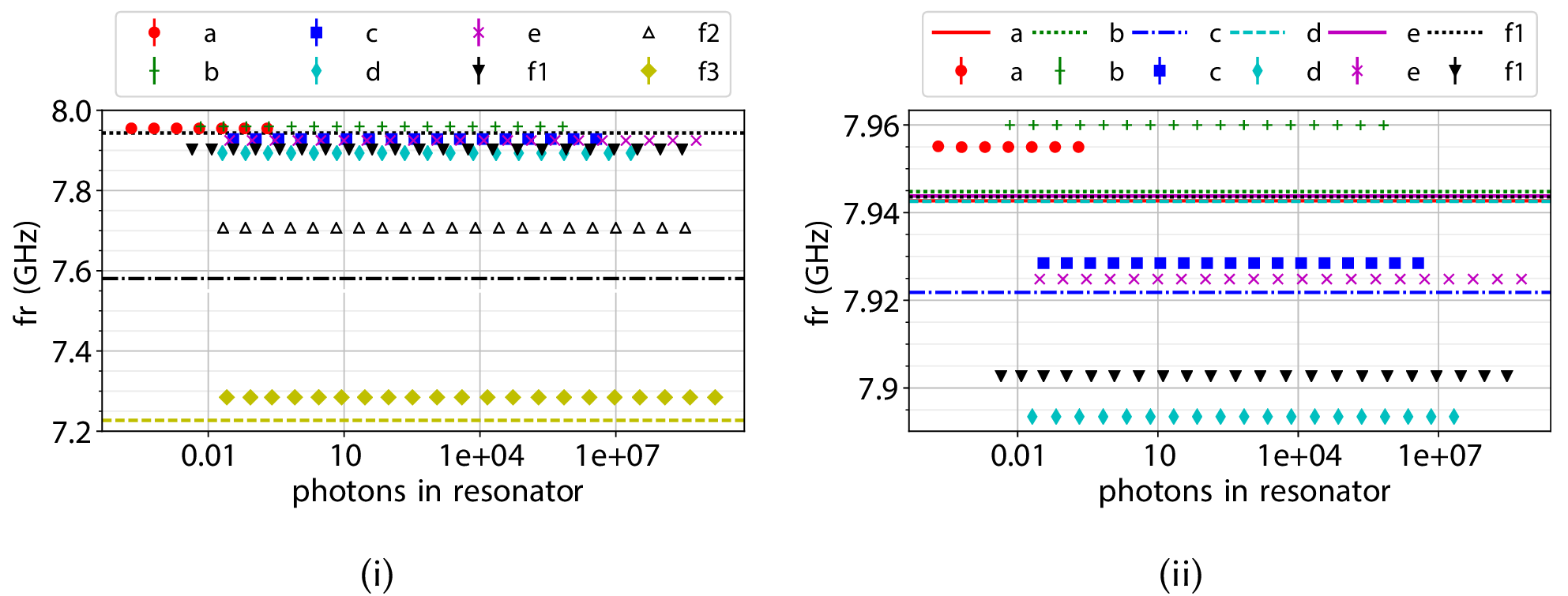}
    \caption{Resonance frequencies for the different configurations (see Fig.~\ref{fig:wg_setups}) in dependence of photon number. The lines depict the simulated data, the points are the measurements. (i) All measured configurations. (ii) Configurations with nominally the same resonance frequency. The deviation of the simulated resonance frequency for configuration (c) can be explained with the closer sapphire substrate, compared to the other setups. The sapphire leads to a higher effective $\epsilon_r$ and thus a lower resonance frequency.}
    \label{fig:fr_meas}
\end{figure}

The measured resonance frequencies are shown in Fig.~\ref{fig:fr_meas}. The additional silicon substrate backing reduces the resonance frequencies of the \SI{7.5}{GHz}(f3) and \SI{8}{GHz}(f2) MSRs, plotted in (i). Except (f1), simulation results accurately predict the resonance frequencies. All the other setups have resonance frequencies in the same range (Fig.~\ref{fig:fr_meas}(ii)), which is predicted by simulations. There are several explanations for the \SI{70}{MHz} deviation of the resonance frequency, which is not seen in the simulation data. One possibility is, a variance in the chip dimension. This would lead to a different effective dielectric constant and thus a lower resonance frequency. Other possibilities include a slight difference between the MSRs or its placement on the substrate. 

There is no dependence of the resonance frequency on the number of circulating photons.

\subsection*{Shift of the resonance frequency of the Al MSR with increasing temperature}

\begin{figure}[ht]
    \centering
    \includegraphics[width = 3.37in]{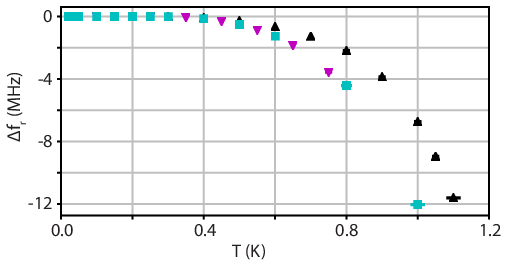}
    \caption{Shift of the resonance frequencies for increasing base temperature of the aluminum MSR. \protect \BlackTriaUp \protect \CyanRect Al MSR in copper waveguide. \protect \MagentaTriaDown Al MSR in aluminum waveguide. }
    \label{fig:Al_fr_temp}
\end{figure}

The shift of the resonance frequency of the three measured aluminum MSRs for increasing base temperature is plotted in Fig.~\ref{fig:Al_fr_temp}. A decrease of the resonance frequency is seen above \SI{500}{mK}. The shift is similar for all three measured samples. The drop in resonance frequency can be explained with an increasing surface inductance over temperature, which originates from an increasing effective penetration depth~\cite{S_gao_physics_2008}. The increase of the penetration depth can be estimated with the Mattis Bardeen theory~\cite{S_mattis_theory_1958}.
\\The results are similar to the one found in~\cite{S_reagor_reaching_2013, S_gao_physics_2008}, where thin aluminum film resonators were measured. There, the frequency shift shows good agreement with the Mattis-Bardeen theory.

\subsection*{Resonance frequency shift of the niobium MSR with increasing temperature - comparison between low and high input powers}

\begin{figure}[ht]
    \centering
    \includegraphics[width = 3.37in]{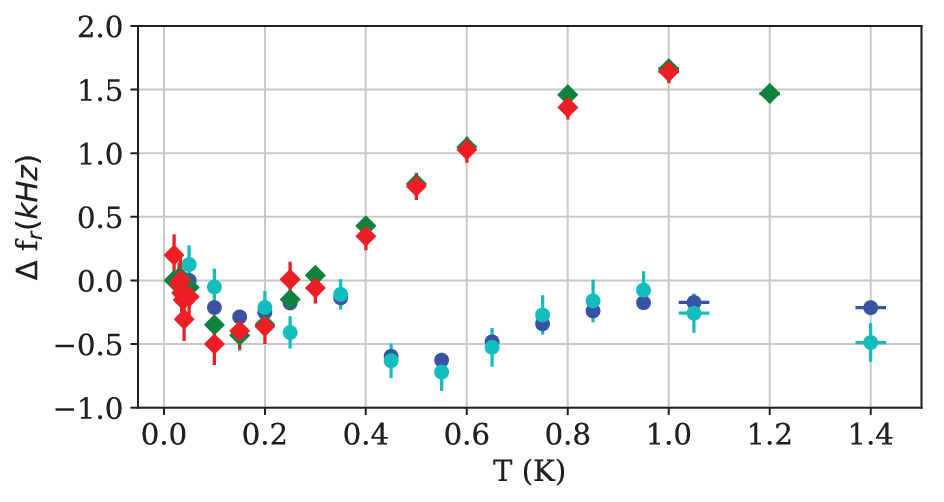}
    \caption{Shift of the resonance frequency of the Nb MSR with increasing temperature, discussed in the main article. High power measurements were taken for the fit in the main article (Fig.~4(b)). \protect \GreenDia $\approx 10^6$ photons circulating in resonator \protect \RedDia single photon limit for MSR in a copper waveguide. \protect \BlueCircle $\approx 10^6$ photons circulating in resonator \protect \CyanCircle  single photon limit for MSR in aluminum waveguide.}
    \label{fig:Nb_fr_temp}
\end{figure}

Fig.~\ref{fig:Nb_fr_temp} compares the shift of the resonance frequency for input powers at the single photon limit to input powers six magnitudes greater. The main difference is the higher noise in the single photon limit leading to increasing uncertainties. Overall, the low and high power measurements show the same temperature dependence. Thus both can be taken to fit $\Delta f_r$ with the same results. Given the lower uncertainties, we took the high power measurements to perform the fit.

In Fig.~\ref{fig:Nb_fr_temp} the measurement results are plotted until \SI{1.4}{K}. For the fit to the MSR in the copper waveguide, the data points above \SI{0.8}{K} are omitted as the behavior above is not well described by the model anymore.

\subsection*{Internal quality factor of the niobium MSR for high excitation powers}

\begin{figure}[ht]
    \centering
    \includegraphics[width = 3.37in]{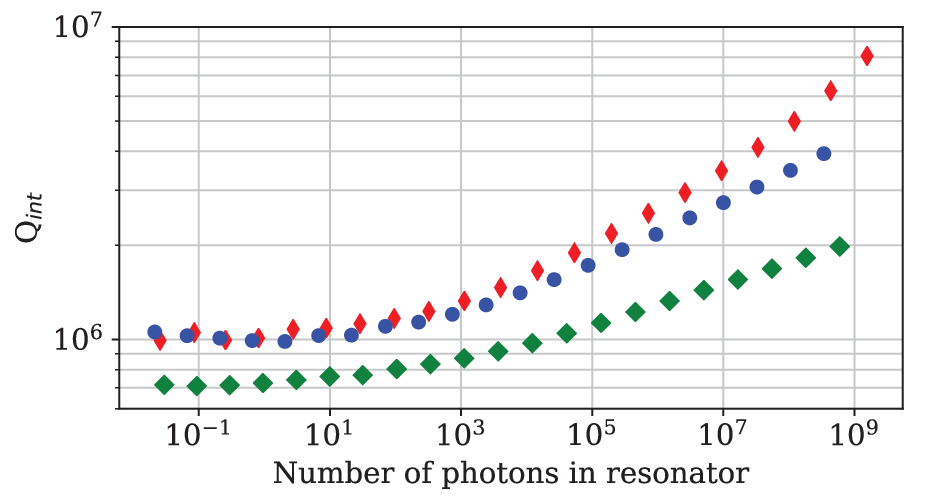}
    \caption{Similar to Fig.~2 in the main article but showing the full measurement range of the niobium MSR. The internal quality factor reaches nearly ten million. \protect \GreenDia Nb MSR in copper waveguide. \protect \BlueCircle \protect \RedDia Nb MSR in aluminum waveguide.}
    \label{fig:Nb_Qi_Pram_full}
\end{figure}

Fig.~\ref{fig:Nb_Qi_Pram_full} shows the internal quality factor of the niobium MSR over the whole measurement range. The best performing niobium MSR showed an internal quality factor of above eight million for high input powers. Due to attenuators in the measurement chain (Fig.~\ref{fig:mm_setup}) higher input powers were not possible. Given the tendency, we would expect an even higher quality factor for higher input powers. The difference between the two MSRs in the aluminum waveguide is probably related to TLS losses. We also find a slightly lower value for the combined loss parameter $k$ of the better performing MSR (Tab.~\ref{tab:fit_res_Nb_TempRamp}). The finite conductivity of the copper is probably the reason for the lower quality factor measured for the MSR in the copper waveguide.

\section*{Fit results}
In this section we provide the fit parameters to the fits, that were discussed in the main part.

The internal quality factor of the MSR changes with temperature. In case of the niobium MSR this is explained with the loss to two level systems (Eq.~2). We fit this model to the measurement data (Fig.~4(a)). TLS also lead to a shift of the resonance frequency, predicted by Eq.~3. The measurement results including the fits are shown in Fig.~4(b). 

In case of the aluminum MSR, an increasing surface resistance also leads to an additional effect on $Q_{\text{int}}$, next to the TLS. Thus we use a combined model of the surface impedance (Eq.~1) and TLS (Eq.~2) for the fit:
\begin{equation}
     \frac{1}{Q_{\text{int}}^{\text{TLS + R}_\text{s}}} = k \tanh{ \left( \frac{h f_r(T)}{2 k_B T} \right) } + \frac{A}{T} exp{ \left( - \frac{\Delta}{k_B T} \right) } + Q_{\text{other}}
     \label{equ:model_alQi_all}
\end{equation}
To fit this model to the change of $Q_{\text{int}}$, the inverse of Eq.~\ref{equ:model_alQi_all} was taken. The fit parameters of all performed fits are listed here. In Tab.~\ref{tab:fit_res_QiAl_TempRamp} the fit results of Eq.~\ref{equ:model_alQi_all} to the measurements of the aluminum MSR (Fig.~3) are given.

\begin{table}[ht]
    \caption{Fit results including fit errors for the measurement of the internal quality factor of the aluminum MSR, for stepwise increasing temperature. The results including the fits are shown in Fig.~3. The corresponding model is given in Eq.~\ref{equ:model_alQi_all}. Given that the the MSRs in the copper waveguide are not limited by TLS related losses, the fit values are not trustworthy. A similar conclusion can be drawn for the MSR in the aluminum waveguide, for $Q_{\text{other}}$. The MSR seems to be either limited by TLS or surface impedance losses, so $Q_{\text{other}}$ is not seen in the measurement.}
    \centering
    \begin{tabular}{l c|c|c|c}
    & & $A$ & $Q_{\text{other}}$ & $k$ \\ \hline
    Al - cu & (\MagentaTriaDown) & \SI{5.6(7)e-4}{} & \SI{4(1)e5}{} & \SI{1.2(6)e-6}{} \\ \hline
    
    Al - cu & (\CyanRect) & \SI{10.2(9)e-4}{} & \SI{0.92(2)e5}{} & \SI{4.46e-1}{} $\pm$ \SI{2.1}{} \\ \hline
    
    Al - al & (\BlackTriaUp) & \SI{6.5(7)e-4}{} & - & \SI{1.6(2)e-6}{} \\

    \end{tabular}

    \label{tab:fit_res_QiAl_TempRamp}
\end{table}

In case of the aluminum MSR in the copper waveguide, $1/k$ can not be taken as a low energy low temperature limit for $Q_{\text{int}}$, as the MSR is limited by other losses. Only the MSR in the aluminum waveguide, being limited by TLS related losses, $1/k =$ \SI{6(1)e5}{}, is in agreement with the measurements. In turn, it was not possible to extract a useful value for $Q_{\text{other}}$, as the MSR was either limited by TLS effects or increasing conductive losses. For the MSRs in the copper waveguide $Q_{\text{other}}$ is in agreement with the measurements. The values obtained for $A$, which refers to the increasing surface impedance, are in the same range for all measurements.

Tab.~\ref{tab:fit_res_Nb_TempRamp} lists the fit results for the niobium MSR. We fitted both, the change of the internal quality factor with temperature (Fig.~4(a)) and the change of the resonance frequency with temperature (Fig.~4(b)).
\begin{table}[ht]
    \caption{Fit results including fit errors for the measurement of the internal quality factor of the niobium MSR, for stepwise increasing temperature. The measurement results including the fits are shown in Fig.~4. The corresponding model for the change of $Q_{\text{int}}$ is given in Eq.~2 and for the change of the resonance frequency in Eq.~3.}
    \centering
    \begin{tabular}{l c|c|c||c}
    \multicolumn{2}{c}{ } & \multicolumn{2}{c}{fit to $Q_{\text{int}} (T)$} & \multicolumn{1}{c}{fit to $\Delta f_r (T)$}\\  
    & & $k$ & $Q_{\text{other}}$ & $k$ \\ \hline
     
    Nb - cu & (\GreenDia) & \SI{10.0(2)e-7}{} & \SI{2.53(4)e6}{} & \SI{9.1(5)e-7}{} \\ \hline
    
    Nb - al & (\BlueCircle) & \SI{7.5(4)e-7}{} & \SI{4.7(5)e6}{} & \SI{5.6(3)e-7}{} \\ \hline
    
    Nb - al & (\RedDia) & \SI{7.9(7)e-7}{} & \SI{7(2)e6}{} & \SI{6.1(4)e-7}{} \\
    
    \end{tabular}
    
    \label{tab:fit_res_Nb_TempRamp}
\end{table}
For the niobium MSR the $k$ can be either determined by fitting the change of $Q_{\text{int}}$ or the change of the resonance frequency. In both cases the value we got are in the same range. Nevertheless, the value obtained fitting the resonance frequency is throughout $10\% - 30\%$ higher, than fitting $Q_{\text{int}}$. The reason is explained in the main article and lies in the frequency distribution of the TLS~\cite{S_pappas_two_2011}. In addition, the $Q_{\text{int}}$ limit given through $k$ is also discussed in the main article. $Q_{\text{other}}$ gives an upper limit on the internal quality factor. The fit values are compatible with the measurements (Fig.~\ref{fig:Nb_Qi_Pram_full}).

\end{CJK*}

\end{document}